\documentclass{ws-ijmpa}
\usepackage[super,compress]{cite}
\usepackage{graphicx}
\usepackage{braket}
\usepackage{bm}% bold math
\usepackage{comment}
\usepackage{here}
\usepackage{multirow}
\usepackage{xcolor}

\begin{document}
\markboth{Teruyuki Kitabayashi}{Primordial black holes and mirror dark matter}

%%%%% Personal Macros %%%%%%%%%%%%%%%%%%%
% Journal macro
\def\Journal#1#2#3#4{{#1} {\bf #2}, #3 (#4)}
% Journal names
\def\AHEP{Advances in High Energy Physics.} 
\def\ARNPS{Annu. Rev. Nucl. Part. Sci.} 
\def\AandA{Astron. Astrophys.} 
\def\ANP{Ann. Phys.}
\def\APJ{Astrophys. J.}
\def\APJS{Astrophys. J. Suppl}
\def\CMP{Commn. Math. Phys.}
\def\COMR{Comptes Rendues}
\def\CPC{Chin. Phys. C}
\def\CQG{Class. Quant. Grav.}
\def\EPJC{Eur. Phys. J. C}
\def\EPJP{Eur. Phys. J. Plus}
\def\FASS{Front. Astron. Space Sci.}
\def\IJMPA{Int. J. Mod. Phys. A}
\def\IJMPD{Int. J. Mod. Phys. D}
\def\IJMPE{Int. J. Mod. Phys. E}
\def\JCAP{J. Cosmol. Astropart. Phys.}
\def\JHEP{J. High Energy Phys.}
\def\JETP{J. Exp. Theor. Phys}
\def\JETPL{JETP. Lett.}
\def\JETPUSSR{JETP (USSR)}
\def\JPG{J. Phys. G} 
\def\JPGNP{J. Phys. G: Nucl. Part. Phys.} 
\def\MPLA{Mod. Phys. Lett. A}
\def\MNRAS{Mon. Not. R. Astron. Soc.}
\def\NIMA{Nucl. Instrum. Meth. A.}
\def\NATU{Nature}
\def\NATULONDON{Nature (London)}
\def\NCA{Nuovo Cimento}
\def\NJP{New. J. Phys.}
\def\NPB{Nucl. Phys. B}
\def\NPBOLD{Nucl. Phys.}
\def\NPBSUPPL{Nucl. Phys. B. Proc. Suppl.}
\def\PDU{Phys. Dark. Univ.}
\def\PLB{{Phys. Lett.} B}
\def\PMCA{PMC Phys. A}
\def\PREP{Phys. Rep.}
\def\PPNP{Prog. Part. Nucl. Phys.}
\def\PLBOLD{Phys. Lett.}
\def\PAN{Phys. Atom. Nucl.}
\def\PRL{Phys. Rev. Lett.}
\def\PRD{Phys. Rev. D}
\def\PRC{Phys. Rev. C}
\def\PR{Phys. Rev.}
\def\PTP{Prog. Theor. Phys.}
\def\PTEP{Prog. Theor. Exp. Phys.}
\def\RMP{Rev. Mod. Phys.}
\def\RPP{Rep. Prog. Phys.}
\def\SJNP{Sov. J. Nucl. Phys.}
\def\SPJETP{Sov. Phys. JETP}
\def\SPP{SciPost Phys.}
\def\SCIENCE{Science}
\def\TNYAS{Trans. New York Acad. Sci.}
\def\ZETP{Zh. Eksp. Teor. Piz.}
\def\ZFPH{Z. fur Physik}
\def\ZPC{Z. Phys. C}
%%%%%%%%%%%%%%%%%%%%%%%%%%%%%%%%%%%%%%%%%%%%%%%%%%%%%%%%%%%%%%%%%%%%%%%%%%%%%%%%%%%%

\catchline{}{}{}{}{}
%
%%%%%%%%%%%%%%%%%%%%%%%%%%%%%%%%%%%%%%%%%%%%%%%%%%%%%%%%%%%%%%%%%%%%

%\preprint{TOKAI-HEP/TH-0505}

\title{Primordial black holes and mirror dark matter}
% Force line breaks with \\

\author{Teruyuki Kitabayashi}

\address{Department of Physics, Tokai University,\\
4-1-1 Kitakaname, Hiratsuka, Kanagawa 259-1292, Japan\\
teruyuki@tokai-u.jp}

\maketitle

\begin{history}
\received{Day Month Year}
\revised{Day Month Year}
\end{history}

\begin{abstract}
If mirror matter exists but cannot comprise all of the dark matter (DM) in the universe, we can expect that the additional DM component may only interact with the other sectors  gravitationally. One of the natural candidate of a gravitationally interacting component is a primordial black hole (PBH). Therefore, if mirror matter exists but cannot comprise all of the DM in the universe, the existence of PBH may be expected as a candidate of the additional DM component. In this case, the remaining DM components may be PBHs or $SU(3)\times SU(2)\times U(1)$ singlet  particles from PBH. We show constraints on PBH with the mirror DM. Particularly, the initial PBH mass is estimated to be $10^{17} \ {\rm g} \lesssim M_{\rm PBH} \lesssim 10^{23} \ {\rm g}$, if the DM comprises mirror baryons and PBHs.
\end{abstract}

\ccode{12.60.-i,95.35.+d,98.80.Cq}

%\tableofcontents

%%----------------------------------------------------------------------------------
\section{Introduction\label{section:introduction}}
%%----------------------------------------------------------------------------------
The existence of dark matter (DM) is one of the most mysterious questions in particle physics and cosmology. The idea of a mirror copy of the Standard Model particles is attractive for  the DM problem \cite{Ciarcellut2010IJMPD,Foot2014IJMPA}, neutron lifetime puzzle \cite{Berezhiani2019EPJC}, and the internal structure of neutron stars \cite{Ciancarella2021PhysDarkUniv}. 

If there are mirror particles under an unbroken symmetry $G'$, which is a copy of the Standard Model gauge group $G=SU(3) \times SU(2) \times U(1)$, the microphysics of the mirror and the Standard Model sector is identical. Although the renormalizable gauge kinetic mixing and Higgs portal interactions are allowed between the two sectors, these couplings could be sufficiently negligible in the early universe \cite{Roux2020PRD}. Thus Standard Model particles interact with their mirror partners only gravitationally, and the mirror baryons $b'$ may be suitable candidates for DM. The prime denotes the mirror sector.

The mirror baryons are an irreproachable candidate for DM \cite{Clarke2017PLB,Zu2021NPB}. However, the other types of DM could also be present. For example, the mirror baryons could not comprise all of the DM in a galaxy \cite{Roux2020PRD}. With an additional DM component, the DM density in the universe becomes 
\begin{equation}
\Omega_{\rm DM} = \Omega_{b'} + \Omega_{\rm add},
\label{Eq:OmegaDM_OmegaBp_OmegaAdd}
\end{equation}
where $\Omega_{b'}$ denotes the mirror baryon relic density and $\Omega_{\rm add}$ includes the contributions of any other possible DM particles except mirror baryons. Since additional DM components should be singlet under the mirror gauge group $G'$, these components may also be singlet under the Standard Model gauge group $G$. Thus, additional DM components may interact with other sectors gravitationally. 

One of the natural candidate of a gravitationally interacting component ($SU(3)\times SU(2)\times U(1)$ singlet object) is a primordial black hole (PBH). A PBH is a black hole that could have been formed in the early universe \cite{Carr2020ARNPS,Carr2021RPP,Green2021JPGNP,Laha2019PRL,Dasgupta2020PRL,Laha2020PRD,Ray2021PRD,Laha2021PLB,Saha2022PRD}. Additionally, because PBHs emit particles via the Hawking radiation induced by gravity \cite{Hawking1975CMP}, heavy particles produced by Hawking radiation of PBHs could also be a possible candidate for the additional DM \cite{Bell1999PRD,Green1999PRD,Khlopov2006CQG,Baumann2007arXiv,Dai2009JCAP,Fujita2014PRD,Allahverdi2018PRD,Lennon2018JCAP,Morrison2019JCAP,Hooper2019JHEP,Masina2020EPJP,Baldes2020JCAP,Bernal2021JCAP,Gondolo2020PRD,Bernal2020arXiv2,Auffinger2020arXiv,Datta2021arXiv,Chaudhuri202011arXiv,Sandick2021PRD,Cheek2022PRD,Baker2022SciPostPhys,Kitabayashi2021IJMPA,Kitabayashi2022PTEP,Kitabayashi2022ArXiv}. 

Therefore, if mirror matter exists but cannot comprise all of the DM in the universe, the existence of PBH may be expected as a candidate of the additional DM components. In this case, the remaining DM components may be PBHs or $SU(3)\times SU(2)\times U(1)$ singlet particles from PBH. 

In this study, we obtain constraints on PBHs with mirror DM. Particularly, we reveal that the initial PBH mass is estimated to be $10^{17} \ {\rm g} \lesssim M_{\rm PBH} \lesssim 10^{23} \ {\rm g}$, if the DM comprises mirror baryons and PBHs in the context of mirror matter. We note that some connections between PBHs and the mirror sector have been explored for the effective number of neutrinos \cite{Bell1999PRD}, a new method for probing the particle spectrum of nature \cite{Baker2022SciPostPhys}, and a microscopic black hole production and evaporation in particle collisions \cite{Dubrovich2021PRD}.

This paper is organized as follows. In Section \ref{sec:mirror}, we set up a mirror DM scenario. Next, Section \ref{sec:PBHandMirrorDM} presents constraints of PBHs with the mirror sector. Finally, Section \ref{sec:summary} summarizes the study.

In this study, we use the natural unit ($c=\hbar=k_{\rm B}=1$).

%%----------------------------------------------------------------------------------
\section{Mirror dark matter \label{sec:mirror}}
%%----------------------------------------------------------------------------------
We use the simplest model of the mirror sector, which maintains exact mirror symmetry at reheating after inflation. In this model, it is natural that there are only ordinary and mirror matters in the universe at the end of reheating. The microphysics of the mirror and the Standard Model sector are identical. Although the two sectors have the same microphysics, the initial conditions in both sectors could differ. All the differences with respect to the mirror and ordinary sectors can be described in terms of only two free parameters \cite{Ciarcellut2010IJMPD,Foot2014IJMPA}. One is the temperature ratio
\begin{equation}
x=\frac{T'}{T}, 
\label{Eq:x}
\end{equation}
and the other is the mirror baryons $b'$ to ordinary baryons $b$ ratio
\begin{equation}
\alpha=\frac{\Omega_{b'}}{\Omega_b}, 
\label{Eq:R}
\end{equation}
where $T$ ($T'$) and $\Omega_b$ ($\Omega_{b'}$) are the radiation (mirror radiation) temperature and density parameter of baryons (mirror baryons), respectively. We assume that the temperature of the mirror and ordinary sectors are not too different, so that two parameters, $x$ and $\alpha$, are well defined \cite{Ciarcelluti2008PRD}.

The mirror and ordinary particles contribute to the total energy density in the universe. The excess of radiative energy density originating from the relativistic particles in the mirror sector is conventionally expressed as the effective number of neutrinos at big bang nucleosynthesis (BBN) and recombination epochs. The upper limits of the observed effective number of neutrinos \cite{Planck2020AA,Cyburt2016RMP,Hufnagel2018JCAP} yield the upper bound of the temperature ratio
\begin{equation}
x \lesssim 0.5.
\end{equation}
We assume that this constraint is always satisfied, so that the unfavorable relativistic energy density at BBN and recombination epochs was safely suppressed. 

Because the mirror baryon density cannot be more than that inferred for DM \cite{Ciarcellut2010IJMPD,Foot2014IJMPA}, we obtain the upper bound of the $b'b$ ratio
\begin{equation}
0 \le \alpha \le \frac{\Omega_{\rm DM}}{\Omega_b} = 5.33,
\end{equation}
where $\Omega_{\rm DM}$ is the relic density of DM, and we use the data by Planck collaboration $\Omega_{b}=0.0490$ and $\Omega_{\rm DM} = 0.261$ \cite{Planck2020AA}. 

We believe that mirror matter exists but cannot comprise all DM. Thus, the cases where $\alpha = 0$ (there are no mirror baryons) and $\alpha = \Omega_{\rm DM}/\Omega_b=5.33$ (all DM is made of mirror baryons) should be excluded. Although we can expect that the mirror baryonic density is approximately equal to the ordinary one \cite{Berezhiani2001PLB,Bento2001PRL,Foot2004PRD}, small $b'b$ ratio $\alpha \lesssim 0.3 - 0.8$ is also suggested \cite{Roux2020PRD}. A conservative upper and lower bounds of the $b'b$ ratio 
\begin{equation}
0.1 \le \alpha \le 5,
\end{equation}
is considered in our analysis. 

%%----------------------------------------------------------------------------------
\section{Constraints on primordial black holes with mirror dark matter \label{sec:PBHandMirrorDM}}
%%----------------------------------------------------------------------------------

%%----------------------------------------------------------------------------------
\subsection{Heavy PBHs}
%%----------------------------------------------------------------------------------
We assume that PBHs are produced in the early universe by large density perturbations generated from inflation \cite{Garcia-Bellido1996PRD,Kawasaki1998PRD,Yokoyama1998PRD,Kawasaki2006PRD,Kawaguchi2008MNRAS,Kohri2008JCAP,Drees2011JCAP,Lin2013PLB,Linde2013PRD}. Additionally, we assume that PBHs form during the radiation-dominated era with a monochromatic mass function for simplicity \footnote{More realistically, PBH mass function should be extended due to the nature of the gravitational collapse at their formation \cite{Green2016PRD,Kuhnel2017PRD,Carr2017PRD,Bellomo2018JCAP,Calcino2018MNRAS,Boudaud2019PRD,DeRocco2019PRD,Laha2019PRL,Arbey2020PRD,Suyama2020PTEP,Laha2020PRD2,Sureda2021MNRAS,Gow2022PRD,Zhou2022MNRAS}.}, and PBHs do not have angular momentum and electric charge.

We introduce the dimensionless parameter
\begin{eqnarray}
\beta = \frac{\rho_{\rm PBH}}{\rho_{\rm rad}},
\label{Eq:beta}
\end{eqnarray}
to represent the initial density of PBHs at the time of their formation.

A black hole loses its mass $M_{\rm BH}$ by generating particles with masses below the Hawking temperature $T_{\rm BH} = 1/(8\pi G M_{\rm BH})$ via Hawking radiation where $G$ is the gravitational constant \cite{Hawking1975CMP}. 

A PBH with an initial mass larger than $\sim 10^{15}$ g (i.e., heavy PBH) can survive in today's universe \cite{Carr2021RPP}. In this case, the relic density of PBHs is obtained as \cite{Carr2016PRD,Carr2021RPP}
\begin{eqnarray}
\Omega_{\rm PBH} = 1.42\times 10^{17}  \left(\frac{\gamma}{0.2}\right)^{1/2}\left(\frac{0.67}{h}\right)^{2} \left(\frac{106.75}{g_\ast}\right)^{1/4} \left(\frac{10^{15} {\rm g}}{M_{\rm PBH}}\right)^{1/2} \beta,
\label{Eq:OmegaPBH_heavy} 
\end{eqnarray}
where $h$ denotes the dimensionless Hubble parameter ($h=0.6766$ \cite{Planck2020AA}), $M_{\rm Pl}=G^{-1/2}$ is the Planck mass, $\gamma \simeq 0.2$ denotes a numerical factor that depends on the details of the gravitational collapse \cite{Carr1975APJ}, $g_\ast$ denotes the relativistic effective degrees of freedom for the radiation energy density, and $M_{\rm PBH}$ is the initial PBH mass.

For $M_{\rm PBH} \gtrsim 10^{15} \ {\rm g}$, all additional DM components can be explained by the PBH relics
\begin{equation}
\Omega_{\rm DM} = \Omega_{b'} + \Omega_{\rm PBH}.
\label{Eq:DM=b'+PBH}
\end{equation}
In terms of the $b'b$ ratio in Eq. (\ref{Eq:R}), the relic density of PBHs is expressed as
\begin{eqnarray}
\Omega_{\rm PBH} = \Omega_{\rm DM} - \alpha\Omega_{b}.
\label{Eq:OmegaPBH_heavy_DM_ab}
\end{eqnarray}
Combining Eqs. (\ref{Eq:OmegaPBH_heavy}) and (\ref{Eq:OmegaPBH_heavy_DM_ab}), we have 
\begin{eqnarray}
\beta = 8.54\times 10^{-18}  (0.261- 0.049 \alpha) \left(\frac{M_{\rm PBH}}{10^{15} {\rm g}}\right)^{1/2},
\end{eqnarray}
for $M_{\rm PBH} \gtrsim 10^{15} \ {\rm g}$ where we use $g_\ast=2 \times 106.75$.

\begin{figure}[t]
\begin{center}
\includegraphics[scale=1.0]{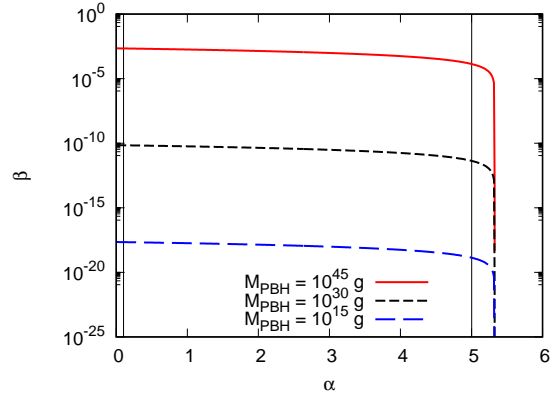}
\caption{Correlation between the initial density of PBHs $\beta$ and the mirror baryon to ordinary baryon ratio $\alpha = \Omega_{b'}/\Omega_b$ for heavy PBHs. Two vertical lines show the conservative lower and upper limits of $\alpha$ ($0.1 \le \alpha \le 5$). The curves for $M_{\rm PBH} = 10^{15}, 10^{30}$ and $10^{45}$g are shown as examples.}
\label{Fig:heavyPBHa} 
\end{center}
\end{figure}

Fig. \ref{Fig:heavyPBHa} shows the correlation between the initial density of PBHs $\beta$ and the mirror to ordinary baryon ratio $\alpha$ for heavy PBHs. Two vertical lines show the conservative lower and upper bounds of $\alpha$ ($0.1 \le \alpha \le 5$). The curves for $M_{\rm PBH} = 10^{15}, 10^{30}$, and $10^{45}$ g are shown as examples. Except for $\alpha \simeq 5.33$, the initial PBH density is well predicted by the $b'b$ ratio for fixed initial PBH mass. Recall that $\alpha = 5.33$ means that all DM is made of mirror baryons. Therefore, $\alpha = 5.33$ should be excluded and we discuss the phenomenology for $0.1 \le \alpha \le 5$. We note that a more stringent bound of $\alpha$ may be $0.1 \le \alpha \le 0.8$ from Ref.\cite{Roux2020PRD}. However, this does not change the conclusion for our numerical estimations, because the $\alpha$ dependence of $\beta$ is moderate except around $\alpha=5.33$ as shown in Fig. \ref{Fig:heavyPBHa}.

\begin{figure}[t]
\begin{center}
\includegraphics[scale=1.0]{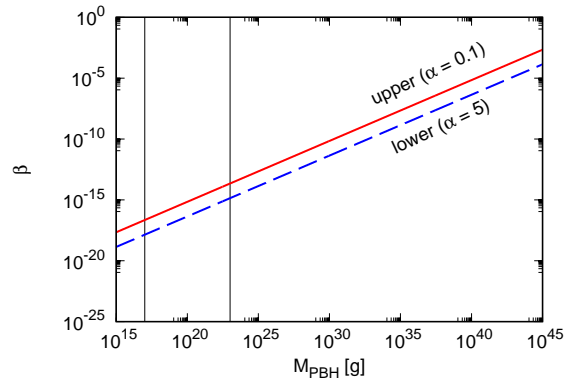}
\caption{The upper bound (curve for $\alpha=0.1$) and the lower bound (curve for $\alpha=5$) of the initial PBH density $\beta$ as a function of initial PBH mass $M_{\rm PBH}$ for heavy PBHs. Two vertical lines show the expected allowed region of the initial PBH mass. }
\label{Fig:heavyPBHb} 
\end{center}
\end{figure}

Fig. \ref{Fig:heavyPBHb} shows the upper bound (curve for $\alpha=0.1$) and the lower bound (curve for $\alpha=5$) of the initial PBH density $\beta$ as a function of initial PBH mass $M_{\rm PBH}$ for $M_{\rm PBH} \ge 10^{15}$ g. 

A summary of the observed upper bound of the initial PBH density is provided by Carr et al., (Fig. 18 in Ref. \cite{Carr2021RPP}. See also Ref. \cite{Carr2020ARNPS}). The predicted $\beta$ should be less or equal than the observed upper bound. We found the predicted lower bound (the line for $\alpha=5$ in Fig. \ref{Fig:heavyPBHb}) for $10^{15} \ {\rm g} \lesssim M_{\rm PBH} \lesssim 10^{17} \ {\rm g}$ and $10^{23} \ {\rm g} \lesssim M_{\rm PBH} \lesssim 10^{45} \ {\rm g}$ are higher than observed upper bound in Ref. \cite{Carr2021RPP}. 

Thus, if both mirror baryons and PBHs are DM in the universe, the allowed region of the initial PBH mass is estimated to be $10^{17} \ {\rm g} \lesssim M_{\rm PBH} \lesssim 10^{23} \ {\rm g}$. In Fig. \ref{Fig:heavyPBHb}, two vertical lines display the allowed region of the initial PBH mass. The corresponding upper (lower) bound of the initial PBH density $\beta$ is predicted to be between $2.2 \times 10^{-17}$ ($1.4 \times 10^{-18}$) for $M_{\rm PBH} \simeq 10^{17}$ g and $2.2 \times 10^{-14}$ ($1.4 \times 10^{-15}$) for $M_{\rm PBH} \simeq 10^{23}$ g.

%--------------------------------------------------------------------
%-------------------------------------------------------------------

%--------------------------------------------------------------------
%-------------------------------------------------------------------

We would like to note that if we know the dark mater fraction $f=\Omega_{\rm PBH}/\Omega_{\rm CDM}$, for example see Ref. \cite{Carr2020ARNPS}, we can predict the initial mass of PBH. Because $\Omega_{\rm CDM}$ is used to predict the initial mass of the PBH in this study, as a result, the argument using $f$ and the argument using $\alpha$ reach similar conclusions. In the future, $\alpha$ could be precisely determined from some arguments, such as neutron lifetimes. If $\alpha$ is not zero and the remaining DM could be the PBH itself or the $SU(3)\times SU(2)\times U(1)$ singlet elementary particle emitted from the PBH, ground-based neutron experiments can provide insight into the mass of the PBH in the universe. Thus, the showing $\alpha-\beta$ relation with $M_{\rm PBH}-\beta$ relation is valuable as well as showing dark matter fraction $f$ in the $M_{\rm PBH}-\beta$ plane in Ref. \cite{Carr2020ARNPS}.

%%----------------------------------------------------------------------------------
\subsection{Light PBHs}
%%----------------------------------------------------------------------------------
A PBH with an initial mass smaller than $10^9$ g (i.e., light PBH) can completely evaporate before BBN \cite{Fujita2014PRD,Kohri1999PRD,Carr2021RPP} and does not affect the successful formation of light elements in BBN. Additionally, because the Hubble parameter $H(t)$ at time $t$ is less than or equal to the Hubble parameter during inflation, we obtain the lower limit of the initial mass of PBH as $M_{\rm PBH} \gtrsim 0.1$ g \cite{Fujita2014PRD}. Thus, we set the mass of light PBHs as $0.1 \ {\rm g} \le M_{\rm PBH} \le 10^9 \ {\rm g}$.

If all PBHs evaporated completely during the radiation-dominant era before BBN, the energy density of particle species $i$ from PBHs evaporation is obtained as \cite{Cheek2022PRD}
\begin{eqnarray}
\Omega_i =  3.553  \left(\frac{\gamma}{0.2}\right)^{1/2} \left(\frac{0.67}{h}\right)^{2} \left(\frac{106.75}{g_\ast}\right)^{1/4} \left(\frac{1{\rm g}}{M_{\rm PBH}}\right)^{3/2} \beta \left(\frac{m_i}{1{\rm GeV}}\right)\mathcal{N}_i, 
\label{Eq:OmegaDM_PBH_i}
\end{eqnarray}
where
\begin{eqnarray}
\mathcal{N}_i = \frac{27g_i}{1024\pi^4(z_i^{\rm ini})^2}\frac{M_{\rm PBH}^2}{M_{\rm Pl}^2} \int_0^{z_i^{\rm ini}}\frac{\Psi_i(z)}{\sum_j g_j\epsilon_j(\tilde{m}_jz)}zdz, \nonumber 
\label{Eq:intN_i}
\end{eqnarray}
with
\begin{eqnarray}
\Psi_i(z) \simeq a_\Psi\{ 1-\left[1+\exp(-b_\Psi\log_{10}z + c_\Psi) \right]^{-d_\Psi} \}, \nonumber
\end{eqnarray}
$z_i^{\rm ini}=m_i/T_{\rm PBH}^{\rm ini}$ (the ratio of particle mass to initial BH temperature), and $\tilde{m}_j = m_j/m_i$. We use the magnitudes of coefficients $a_\Psi, b_\Psi, c_\Psi$, and $d_\Psi$  in Ref. \cite{Cheek2022PRD}.

For $0.1 \  {\rm g} \le M_{\rm PBH} \le 10^9  \ {\rm g}$, the heavy particles produced by Hawking radiation of PBHs can be an additional DM component:
\begin{equation}
\Omega_{\rm DM} = \Omega_{b'} + \Omega_{b'}^{\rm PBH} + \Omega_\chi^{\rm PBH},
\label{Eq:DM=b'+b'PBH+XPBH}
\end{equation}
where $\Omega_{b'}^{\rm PBH}$ and $\Omega_\chi^{\rm PBH}$ denote the relic density of the mirror baryons $b'$ and the heavy particle $\chi$ emitted by Hawking radiation, respectively. We assume that the heavy particle $\chi$ interacts gravitationally with other sectors. In terms of the $b'b$ ratio, the relic density of the heavy particle is expressed as
\begin{eqnarray}
\Omega_\chi^{\rm PBH} = \Omega_{\rm DM} - \alpha\Omega_{b},
\label{Eq:OmegaChi_light_DM_ab}
\end{eqnarray}
where $\alpha\Omega_{b}=\Omega_{b'} + \Omega_{b'}^{\rm PBH}$. Notably, knowledge of the individual relic densities of initially existing mirror baryons $\Omega_{b'}$ and emitted mirror baryons $ \Omega_{b'}^{\rm PBH}$ are not necessary for our analysis, because both the initial and the emitted mirror baryons contribute to $\Omega_{b'}$ in Eq. (\ref{Eq:OmegaDM_OmegaBp_OmegaAdd}). 

In Eq. (\ref{Eq:DM=b'+b'PBH+XPBH}), we ignore the possibility that heavy particle $\chi$ already existed before PBH evaporation, because we assume unbroken mirror symmetry at the end of reheating. If we include this possibility, we have
\begin{equation}
\Omega_{\rm DM} = \Omega_{b'} + \Omega_{b'}^{\rm PBH} + \Omega_\chi^{\rm PBH}+\Omega_\chi,
\end{equation}
where $\Omega_\chi$ denotes the relic density of the initially existing heavy particle $\chi$. The nonvanishing $\Omega_\chi$ may reduce the predictability of the simplest mirror matter model. We ignore the possibility of initially existing heavy particle $\chi$ from the standpoints of the unbroken mirror symmetry and the keeping predictability. Combining Eqs. (\ref{Eq:OmegaDM_PBH_i}) and (\ref{Eq:OmegaChi_light_DM_ab}), we have
\begin{eqnarray}
\beta =  0.341 (0.261 - 0.049\alpha) \left(\frac{M_{\rm PBH}}{1{\rm g}}\right)^{3/2}  \left(\frac{1{\rm GeV}}{m_\chi}\right)\mathcal{N}_\chi^{-1}. 
\end{eqnarray}
%

%--------------------------------------------------------------------
\begin{figure}[t]
\begin{center}
\includegraphics[scale=1.0]{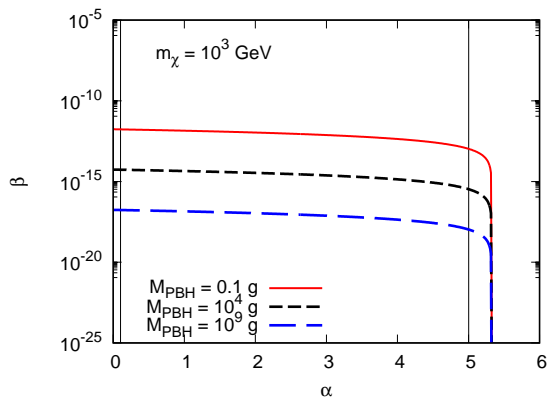}
\caption{Correlation between the initial density of PBHs $\beta$ and the mirror baryon to ordinary baryon ratio $\alpha = \Omega_{b'}/\Omega_b$ for light PBHs. Two vertical lines show the conservative lower and upper limits of $\alpha$ ($0.1 \le \alpha \le 5$). The curves for $M_{\rm PBH} = 0.1, 10^4$ and $10^9$g, and $m_\chi = 10^3$GeV are shown as examples.}
\label{Fig:lightPBHa} 
\end{center}
\end{figure}
%-------------------------------------------------------------------
%--------------------------------------------------------------------
\begin{figure}[t]
\begin{center}
\includegraphics[scale=1.0]{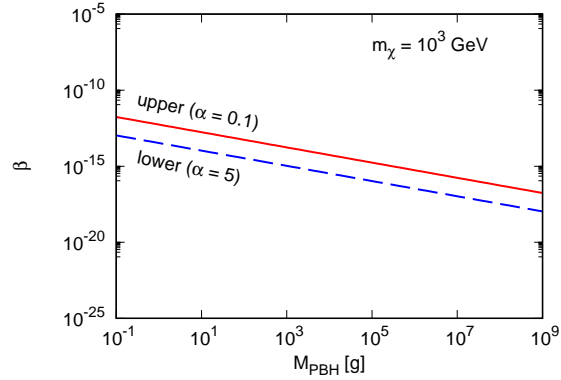}
\caption{The upper bound (curve for $\alpha=0.1$) and the lower bound (curve for $\alpha=5$) of the initial PBH density $\beta$ as a function of initial PBH mass $M_{\rm PBH}$ for light PBHs. The curves for $m_\chi = 10^3$ GeV are shown as examples.}
\label{Fig:lightPBHb} 
\end{center}
\end{figure}
%-------------------------------------------------------------------
%--------------------------------------------------------------------
\begin{figure}[t]
\begin{center}
\includegraphics[scale=1.0]{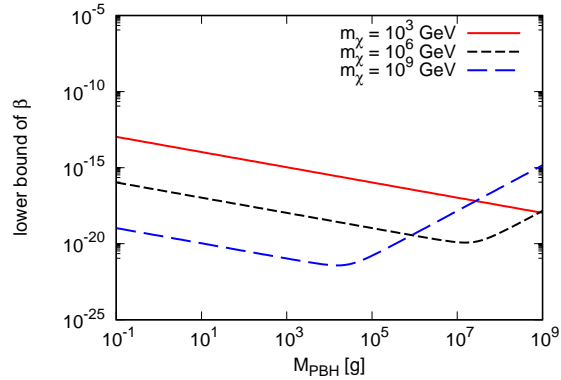}
\caption{The lower bound of the initial PBH density $\beta$ as a function of initial PBH mass $M_{\rm PBH}$ for light PBHs. The curves for $m_\chi = 10^3$, $10^6$, and $10^9$GeV are shown as examples.}
\label{Fig:lightPBHc} 
\end{center}
\end{figure}
%-------------------------------------------------------------------

We set the range of heavy particle mass to be $10^{-2} \ {\rm GeV} \le m_\chi \le 10^9 \ {\rm GeV}$ for illustration. The lower mass $m_\chi = 10^{-2}$ GeV is used to avoid the warm DM consideration with PBHs \cite{Fujita2014PRD,Masina2020EPJP,Cheek2022PRD}. Conversely, because the semiclassical approximation used by Hawking to derive the evaporation spectrum fails at the Planck scale \cite{Dambrosio2021PRD,Soltani2021PRD}, the upper mass should be lighter than the Planck mass, $m_\chi < M_{\rm Pl}$. The upper mass $m_\chi = 10^9$ GeV is sufficient to illustrate a sample case.

We assume that emitted heavy particle $\chi$ is a scalar particle. According to Cheek et al., the relic abundance of emitted particles depends on their spin, varying in $\sim 2$ orders of magnitude between scalar and spin-2 particles in the case of Schwarzschild PBH \cite{Cheek2022PRD}. Although this variation of $\sim 2$ orders of magnitude may have a substantial effect in some situations, this spin effect does not make a significant difference in the conclusion of this paper. 

Fig. \ref{Fig:lightPBHa} shows the correlation between the initial density of PBHs $\beta$ and the mirror to ordinary baryon ratio $\alpha$ for light PBHs. Two vertical lines show the conservative lower and upper limits of $\alpha$ ($0.1 \le \alpha \le 5$). The curves for $M_{\rm PBH} = 0.1, 10^4$, and $10^9$ g, and $m_\chi = 10^3$ GeV are shown as examples. As with the heavy PBH case, except for $\alpha \simeq 5.33$, the initial PBH density is well predicted by the $b'b$ ratio for the fixed initial PBH and the heavy particle masses. We discuss the phenomenology for $0.1 \le \alpha \le 5$.

Fig. \ref{Fig:lightPBHb} shows that the upper bound (curve for $\alpha=0.1$) and the lower bound (curve for $\alpha=5$) of the initial PBH density $\beta$ as a function of initial PBH mass $M_{\rm PBH}$ for $0.1 \ {\rm g} \le  M_{\rm PBH} \le 10^9$ g. The curves for $m_\chi = 10^3$ GeV are shown as examples. 

Unlike the case of heavy PBHs, the initial PBH density depends on three parameters ($\alpha$, $M_{\rm PBH}$, and $m_\chi$). Although the predictive power for the light PBH is less than heavy PBHs, it may be a major advance to demonstrate that a lower bound on the initial density of light PBHs can be obtained under mirror symmetry, because we have minimal knowledge of the observed lower bound of the initial density for light PBHs \cite{Carr2020ARNPS,Carr2021RPP,Green2021JPGNP}. For example, Fig. \ref{Fig:lightPBHc} shows the lower bound of the initial PBH density $\beta$ as a function of initial PBH mass $M_{\rm PBH}$ for light PBHs. The curves for $m_\chi = 10^3$, $10^6$, and $10^9$ GeV are shown as examples. Because $T_{\rm BH} \propto 1/M_{\rm BH}$, heavier initial PBHs produce fewer heavy particles, and leading heavy particle production can begin after crossing $T_{\rm BH} \simeq m_\chi$. Thus, two curves go up at a point for heavy $\chi$ in Fig. \ref{Fig:lightPBHc}. 

Finally, we comment on the PBHs with middle mass  ($10^9 \ {\rm g} \le M_{\rm PBH} \le 10^{15}$ g). Because the middle PBHs may evaporate during BBN, an in-depth analysis is required if we include the middle PBH in our analysis. We want to omit the case of the middle PBHs in this study.

%%----------------------------------------------------------------------------------
\section{Summary\label{sec:summary}}
%%----------------------------------------------------------------------------------
If mirror matter exists but cannot comprise all of the DM in the universe, the additional DM component may interact with the other sectors only gravitationally. One of the natural candidate of a gravitationally interacting component is a PBH. Consequently, if mirror matter exists but cannot comprise all of the DM in the universe, the existence of PBH may be expected as a candidate of the additional DM. In this case, the remaining DM components may be PBHs or $SU(3)\times SU(2)\times U(1)$ singlet particles from PBH. Because the cases of $\alpha =\Omega_{b'}/\Omega_b= 0$ (no mirror baryons) and $\alpha = \Omega_{\rm DM}/\Omega_b=5.33$ (all DM is made of mirror baryons) should be excluded, a conservative upper and lower bounds of the $b'b$ ratio $0.1 \le \alpha \le 5$ has been considered in our analysis. 

If PBHs are heavy ($M_{\rm PBH} \ge 10^{15}$ g), the DM in the universe today may comprise the mirror baryons and the PBHs. In this case, we have found that the allowed region of the initial PBH mass is estimated to be $10^{17} \ {\rm g} \lesssim M_{\rm PBH} \lesssim 10^{23} \ {\rm g}$. The corresponding upper (lower) bound of the initial PBH density $\beta$ may be between $2.2 \times 10^{-17}$ ($1.4 \times 10^{-18}$) for $M_{\rm PBH} \simeq 10^{17}$ g and $2.2 \times 10^{-14}$ ($1.4 \times 10^{-15}$) for $M_{\rm PBH} \simeq 10^{23}$ g. Conversely, if PBHs are light ($0.1 \ {\rm g} \le M_{\rm PBH} \le 10^9$ g), the DM may be made of mirror baryons and the heavy particles emitted from PBHs. For light PBHs, we have shown the lower bound of the initial PBH density $\beta$ as a function of initial PBH mass $M_{\rm PBH}$ for some specific mass of the heavy particles.

\vspace{3mm}
%%%%%%%%%%%%%%%%%%%%%%%%%%%%%%%%%%%%%%%%%%%%%%%%%%%%%%%%%%%%%%%%%%%%%%%%%%%%%%%
%\noindent
%\centerline{\small \bf ACKNOWLEGMENTS}
%%%%%%%%%%%%%%%%%%%%%%%%%%%%%%%%%%%%%%%%%%%%%%%%%%%%%%%%%%%%%%%%%%%%%%%%%%%%%%%

%%%%%%%%%%%%%%%%%%%%%%%%%%%%%%%%%%%%%%%%%%%%%%%%%%%%%%%%%%%%%%%%%%%%%%%%%%%%%%%%% The Appendices part is started with the command \appendix;
%\appendix

%% References

%%References
%%--------------------------------
%\bibliographystyle{elsarticle-num}



%\begin{thebibliography}{000} %for 3 digits
%\begin{thebibliography}{00}  %for 2 digits

\end{document}